\documentclass[aps,prb,reprint,twocolumn,superscriptaddress,showpacs,floatfix]{revtex4-1}

\usepackage{graphicx}
\usepackage{epsfig}
\usepackage{amsmath}
\usepackage{exscale}
\usepackage{array}
\usepackage{units}
\usepackage{amssymb}
\usepackage{setspace}
\usepackage{bbm}
\usepackage{color}
\usepackage{algorithm}
\usepackage{algpseudocode}
\usepackage{color}
\usepackage{enumitem}
\definecolor{lightgrey}{rgb}{0.87,0.87,0.87}

\begin{document}
\title{Generalized self-energy embedding theory}

\author{Tran Nguyen Lan}
\email{latran@umich.edu}
\thanks{Corresponding author}
\altaffiliation{On leave from: Ho Chi Minh City Institute of Physics, VAST, Ho Chi Minh City, Vietnam}
\affiliation{Department of Chemistry, University of Michigan, Ann Arbor, Michigan, 48109, USA}
\affiliation{Department of Physics, University of Michigan, Ann Arbor, Michigan, 48109, USA}
\author{Dominika Zgid}
\email{zgid@umich.edu}
\thanks{Corresponding author}
\affiliation{Department of Chemistry, University of Michigan, Ann Arbor, Michigan, 48109, USA}

\begin{abstract}
{\it Ab initio} quantum chemistry calculations for systems with large active spaces are notoriously difficult and cannot be successfully tackled by standard methods. In this letter, we generalize a Green's function QM/QM embedding method called self-energy embedding theory (SEET) that has the potential to be successfully employed to treat large active spaces. In generalized SEET, active orbitals are grouped into intersecting groups of few orbitals allowing us to perform multiple parallel calculations yielding results comparable to the full active space treatment. We examine generalized SEET on a series of examples and discuss a hierarchy of systematically improvable approximations.  
\end{abstract}

\maketitle

At present, in quantum chemistry there is no established {\it ab initio} method that could treat both strongly correlated molecules and solids while remaining computationally affordable and quantitatively accurate. Methods such as complete active space self-consistent field (CASSCF) \cite{roos1987complete} or complete active space second-order perturbation theory (CASPT2)\cite{Kerstin:jpc/94/5483,Ghigo:cpl/396/142} can treat easily strongly correlated molecules with up to 16 electrons in 16 strongly correlated orbitals but when generalized to solids the strongly correlated orbitals from every cell  add up to a huge overall number making such calculations impossible.
However, due to the experimental progress in solid state chemistry, more than ever the development of a general method capable of treating both strongly correlated molecules and solids while remaining computationally affordable and quantitatively accurate is desired.

The QM/QM embedding methods such as dynamical mean field theory (DMFT)~\cite{Georges96,Georges92,Georges04,savrasov_RevModPhys06,GW_review_werner2016}, density matrix embedding theory (DMET)~\cite{dmet_knizia12,dmet_knizia13}, and self-energy embedding theory (SEET)~\cite{Zgid15,Tran15b,Tran16,zgid_njp17} offer a viable route of generalizing its molecular versions to solids. However, to yield accurate results, molecular versions of these methods must be extensively tested and all the possible sources of inaccuracies must be removed or estimated to deliver systematically improvable and highly accurate answers. 

In this letter, we focus on generalizing the functional form of SEET to successfully overcome the drawbacks of its original formulation. 
SEET is written in the Green's function language providing access not only to total energies but also to photoelectron and angular momentum resolved (ARPES) spectra as well as thermodynamic quantities.  SEET is designed to provide a Green's function functional $\Phi_{SEET}$ that approximates the exact Luttinger-Ward functional $\Phi_{LW}$~\cite{Luttinger60} 
\begin{equation}
\Phi_{SEET}=\Phi^{tot}_{weak}+\sum_{i}(\Phi^{A_i}_{strong}-\Phi^{A_i}_{weak})
\end{equation}
by evaluating $\Phi^{tot}_{weak}$ with a low cost method for all the orbitals present in the system and then selectively improving it by evaluating $\Phi_{strong}^{A_i}$ with a non-perturbative method capable of illustrating strong correlation for $p$ non-intersecting subsets $A_i$ of $l$ strongly correlated orbitals where $p\times l=N$, where $N$ is the total number of strongly correlated orbitals and $M$ is the total number of orbitals in the system and $M\ge N$. 
The $\Phi^{A_i}_{weak}$ part is introduced to remove the double counting of electron correlation in the orbital subsets $A_i$. 

\begin{figure} [htb]
  \includegraphics[width=\columnwidth]{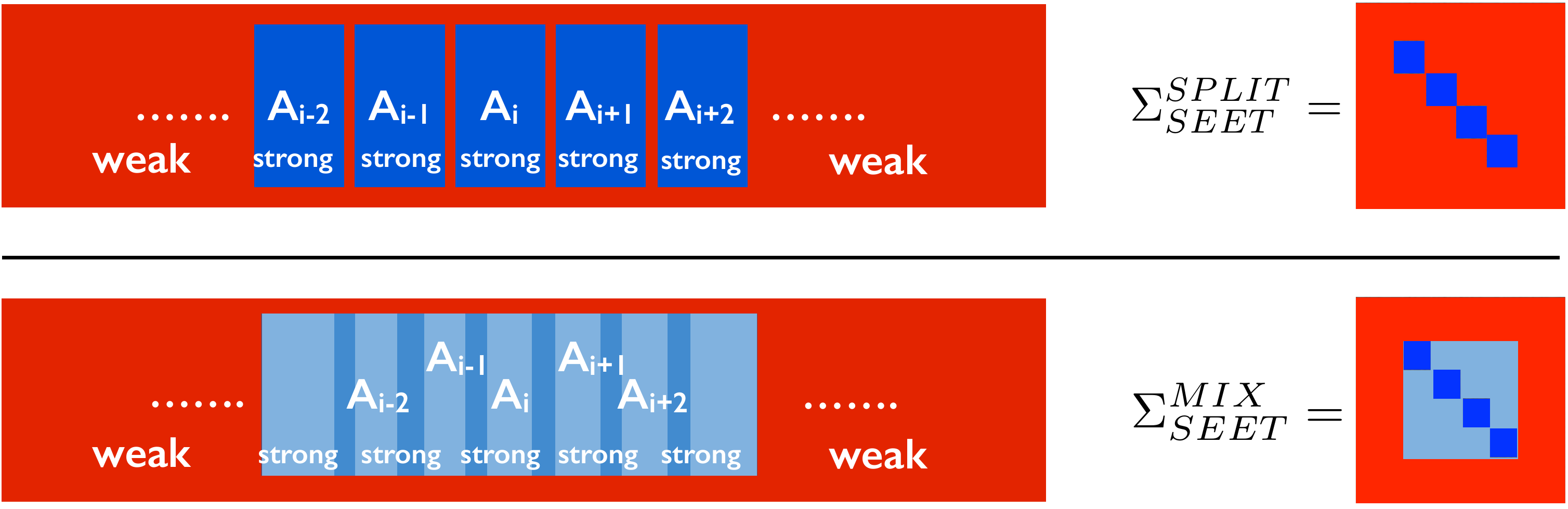} 
  \caption{Left Upper Panel: A system with $M$ orbitals and $p$ non-intersecting groups $A_i$ of $l$ strongly correlated orbitals. Right Upper Panel: The self-energy matrix resulting from SEET treatment of non-intersecting groups of strongly correlated orbitals. Left Lower Panel: A system with $M$ orbitals and intersecting groups $A_i$ of $l$ strongly correlated orbitals. Right Lower Panel: The self-energy matrix resulting from SEET treatment of intersecting groups of strongly correlated orbitals. In both cases strongly correlated orbitals are treated by a non-perturbative, expensive, and accurate method while all the remaining orbitals of the problem are treated by a cheap approximate method.}
  \label{fig:seet_split}
\end{figure}

\begin{figure} [htb]
  \includegraphics[width=\columnwidth]{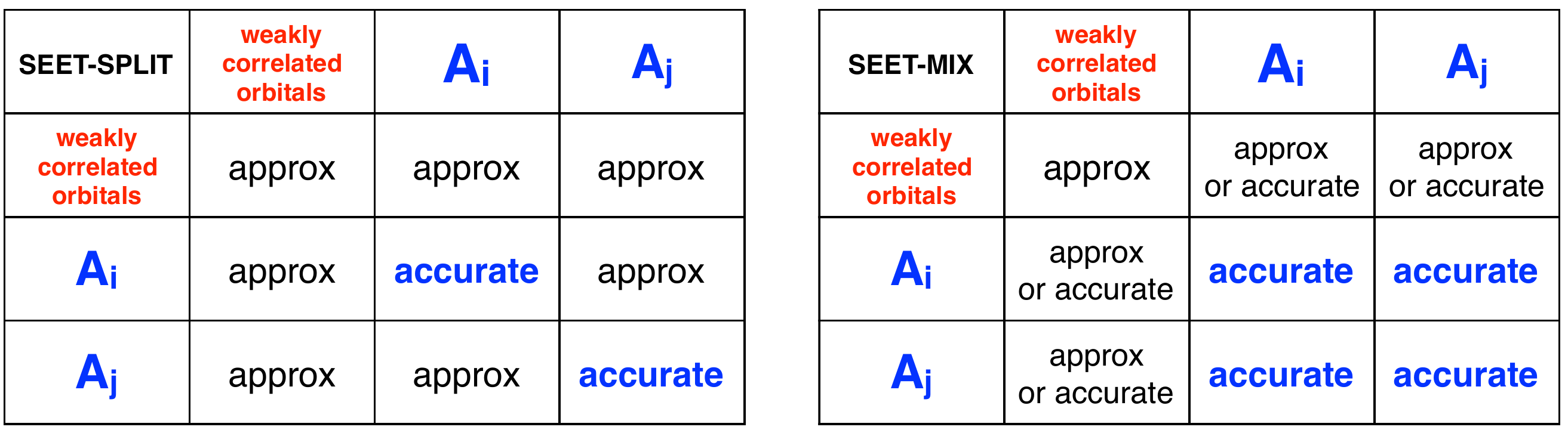} 
  \caption{The treatment of self-energy elements within and between different orbitals groups in SEET with non-intersecting (SEET-split) and intersecting groups (SEET-mix) of orbitals. $A_i$ and $A_j$ are groups of strongly correlated orbitals.}
  \label{fig:interactions}
\end{figure}

In Fig.~\ref{fig:seet_split}, we illustrate the SEET scheme with different orbitals groups and additionally we list the self-energy associated with a SEET functional defined as 
$\Sigma_{ij}=\frac{\partial {\Phi_{SEET}}}{{\partial G_{ij}}}$
where $G$ is the system's Green's function.
Note that $\Phi^{A_i}_{strong}$ and consequently $[\Sigma_{strong}]^{A_i}$ are calculated from appropriate Anderson impurity models (AIMs) which are auxiliary systems used to model the strongly correlated electrons embedded in the field coming from all the other electrons, for details see Ref.~\onlinecite{Tran16} and \onlinecite{zgid_njp17}.
Thus, in the original formulation of SEET only the self-energy elements within each strongly correlated group are treated with an accurate method. The elements of the self-energy between two orbitals belonging to different strongly correlated groups or a strongly and weakly correlated orbital are evaluated at an approximate level. The self-energy of weakly correlated orbitals is also treated by an approximate and cheap method. For details see Fig.~\ref{fig:interactions}.
Moreover, from Fig.~\ref{fig:seet_split} and the form of the original SEET self-energy, it is evident that for cases in which the number of strongly correlated orbitals is increasing and the size of the orbital group $A_i$ remains constant larger and larger part of the self-energy matrix is recovered only by the low cost, approximate method suitable for illustrating weak correlations.
Ultimately such a description may lead to an accuracy loss since only a small part of the self-energy matrix is recovered at an accurate, costly, and  non-perturbative level. 

\begin{figure} [h]
  \includegraphics[width=10.0cm,height=7cm]{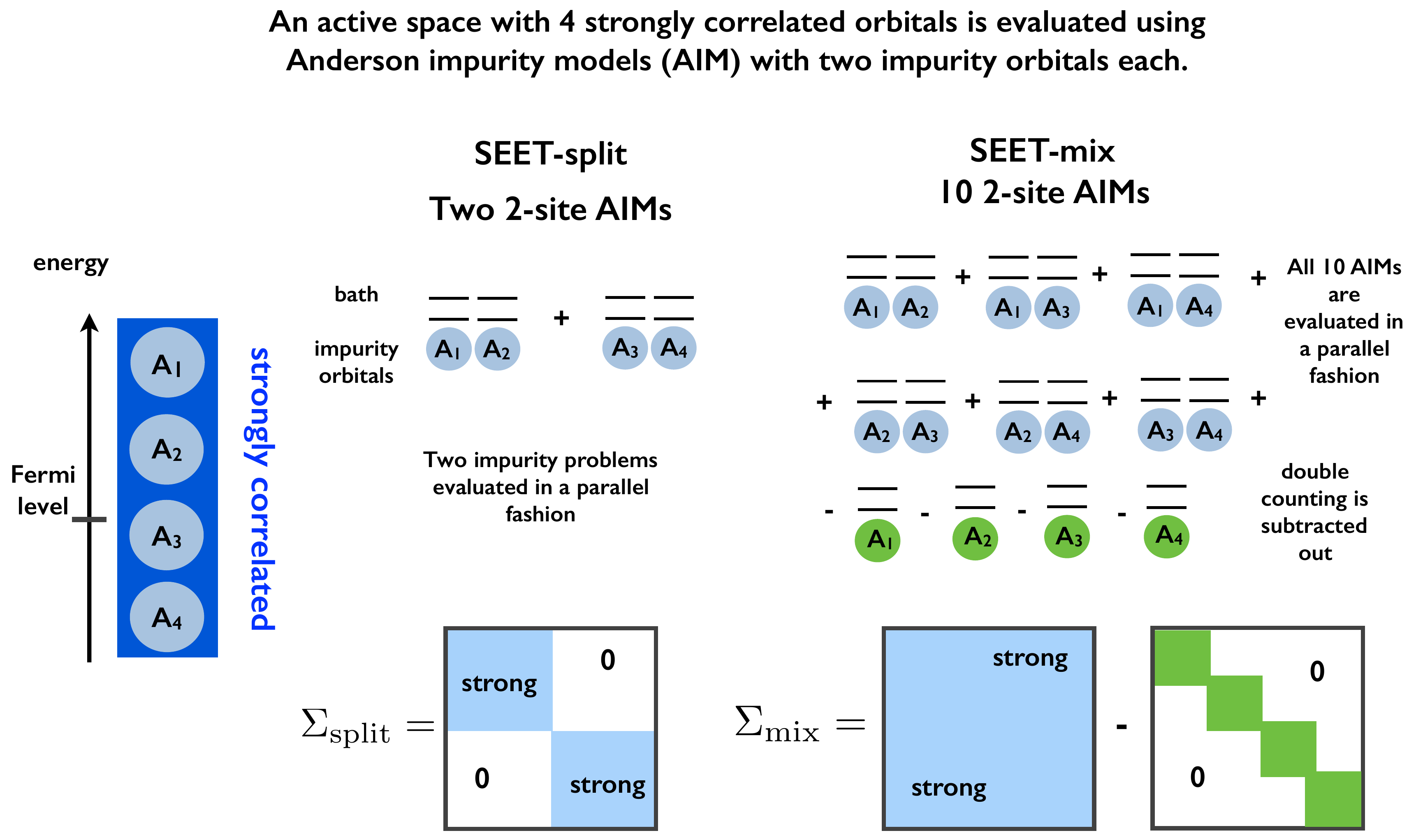}
  \caption{\normalsize A simple schematics for evaluating the self-energy matrix in the SEET-mix approach.}
  \label{fig:sigma_mix_aims_scheme}
\end{figure}

In the generalized formulation of SEET, we overcome this difficulty by producing the self-energy matrix where all the elements between all strongly correlated orbitals (the ones within the groups and between the groups)  are described at the expensive and accurate level, thus creating a better approximation to the exact self-energy, for schematics see  Fig.~\ref{fig:seet_split} and Fig.~\ref{fig:interactions}.
This is achieved by grouping strongly correlated orbitals (or any orbitals) into the intersecting groups. Since such a SEET formulation leads to the double counting of electron correlation coming from the intersecting orbital groups a procedure to  subtract out these double counting contributions must be carried out.
Let us illustrate this procedure with an example. Let us assume that we have a system with 4 orbitals in total and we are able to evaluate an accurate self-energy with an expensive method for two orbitals only. Then we can create $\binom{4}{2}$  groups containing two orbitals $[(1,2),(1,3),(1,4),(2,3),(2,4),(3,4)]$. Consequently, the self-energy for orbitals $1,2,3$,and $4$ is evaluated multiple times (here 3 times). We need to evaluate the self-energy for these single orbitals and subtract it from the self-energy matrix as follows
\begin{eqnarray} \nonumber
\Sigma^{MIX}_{strong}= &\Sigma^{(1,2)}+\Sigma^{(1,3)}+\Sigma^{(1,4)}+\Sigma^{(2,3)}+\Sigma^{(2,4)}+\Sigma^{(3,4)}\\  
&-2\times[\Sigma^{(1)}+\Sigma^{(2)}+\Sigma^{(3)}+\Sigma^{(4)}]. 
\end{eqnarray}
A detailed discussion of the above example is in Fig.~\ref{fig:sigma_mix_aims_scheme} of the supporting information (SI).
The above considerations can be generalized to arbitrary number of strongly correlated orbitals $N$ from which we choose groups of $K$ orbitals. The general form of a SEET functional can be written as
\begin{eqnarray}\label{eq:seet_mix_func} 
\Phi_{SEET}=&\Phi^{tot}_{weak}+(\Phi^{A_i}_{strong}-\Phi^{A_i}_{weak}) \nonumber \\ 
&\pm\sum_{k=K-1}^{k=1}\sum^{\binom{N}{k}}_{i}(\Phi^{B^k_i}_{strong}-\Phi^{B^k_i}_{weak}),
\end{eqnarray}
where the contributions with $\pm$ signs are used to account correctly for the possible double counting of self-energy matrix elements. 
The self-energy matrix is a functional derivative of the above functional with respect to Green's function. 
Note, that all the intersecting groups of strongly correlated orbitals $A_i,\dots,B^{k}_i$ $\forall i,k$  can be treated simultaneously using parallel computing. 
While theoretically it is necessary to include all the contributions to the functional and evaluate $\binom{N}{K}$,$\binom{N}{K-1}$,$\dots$,N  impurity problems necessary to obtain the self-energy, in practice certain groups can be easily excluded. For molecules (or any system with a sizable gap), it is not necessary to create groups of exclusively occupied or virtual orbitals since the self-energy obtained from these groups is close to zero. Our observations indicate that one can further restrict the number of possible orbital groups severely and get very good results if one includes equal number of bonding and anti-bonding orbitals in each of the orbital groups. Another possibility of restricting the number of intersecting groups is to use localized orbitals since only the overlapping orbitals should be within one group. Note that the selection of orbital groups can be done either in the natural or molecular orbital basis or based on spatial distribution of the orbitals in a local orbital basis.
\begin{figure} [htb]
  \includegraphics[width=\columnwidth]{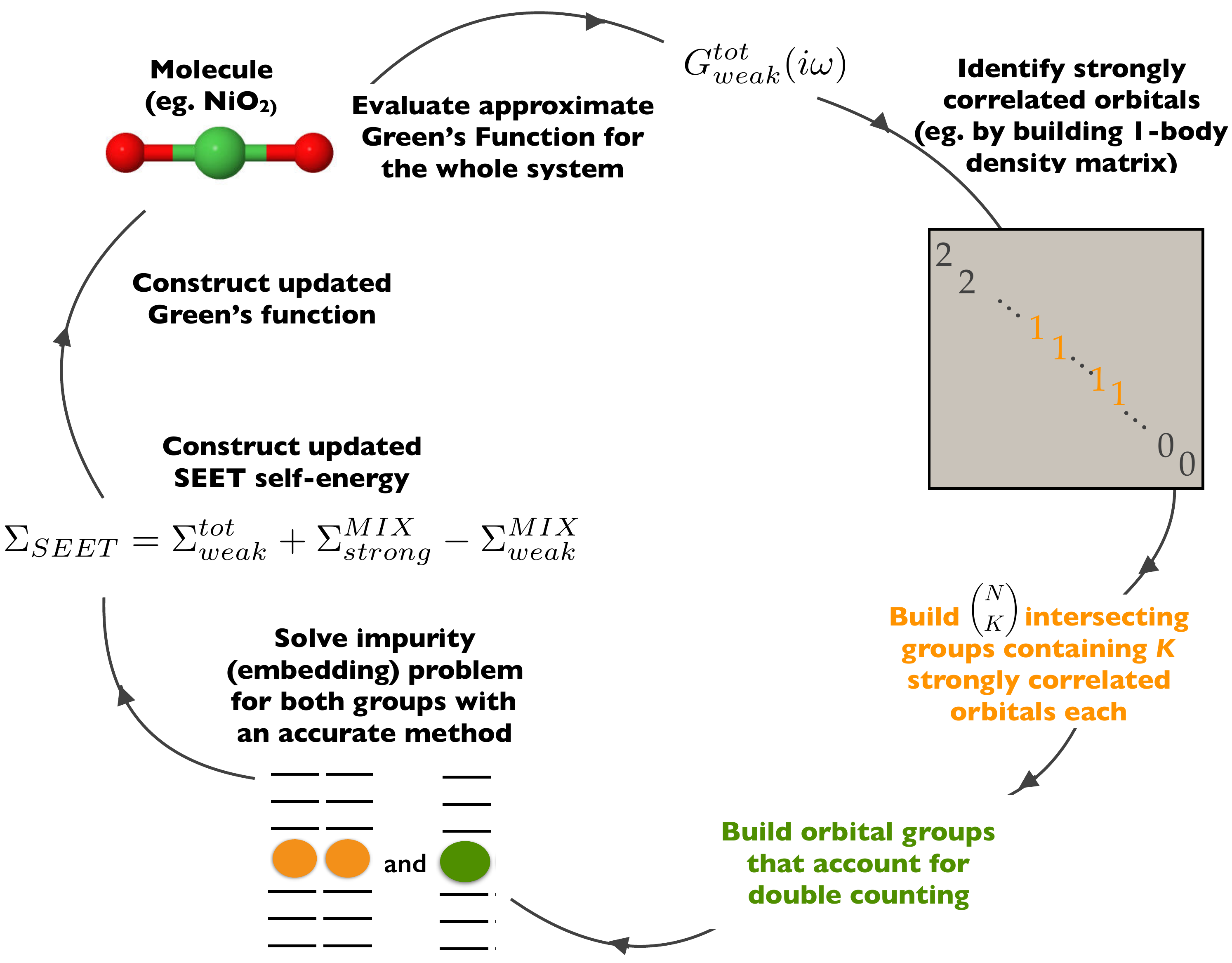} 
  \caption{SEET-mix algorithmic scheme.}
  \label{fig:seet_mix_scheme}
\end{figure}

We list major algorithmic steps necessary to perform the generalized SEET algorithm in Fig.~\ref{fig:seet_mix_scheme}. The whole approximate system Green's function can be evaluated at the Hartree-Fock (HF), Green's function second order (GF2)~\cite{Dahlen05,Zgid14,Phillips15,Kananenka15,Kananenka16,Rusakov16,Welden16}, or GW~\cite{Hedin65,GW100} level. The selection of the strongly correlated orbitals can be done based on the partial occupations of the 1-body density matrix produced in GF2~\cite{Tran16} or GW. In HF, the shape of molecular orbitals and the chemical intuition can be used to guide the selection. Subsequently, based on the computational cost of the accurate solver capable of treating AIMs in the embedding construction, we choose the number of strongly correlated orbitals $K$ that is assigned to each group. 
In our work, the full configuration interaction (FCI) or its truncated variants~\cite{Zgid11,Zgid12} are used as solvers for AIMs.
Since the strongly correlated orbital groups are intersecting (multiple groups may contain the same orbital), we need to ensure that the resulting elements of self-energy are not included multiple times and double counted. 
Consequently, AIMs with impurity orbitals belonging to the intersects are constructed and resulting self-energies are subtracted out from the total sum of self-energies coming from multiple impurities~\footnote{The procedure of eliminating the double counting coming from the common orbitals by subtracting self-energies evaluated by building AIMs made out of orbitals belonging to the intersect rather than simply neglecting repeating elements is absolutely crucial to preserve the conservation laws of the resulting Green's functions.}. For a general case this is illustrated at the functional level in Eq.~\ref{eq:seet_mix_func}. Finally, all the elements of self-energy coming both from the weakly and strongly correlated orbitals are collected and assembled in the total self-energy matrix used to build the Green's function that can be used in subsequent iterations. 

We denote the SEET method where the strongly correlated orbitals are separated into non-intersecting groups as SEET(method$_\text{strong}$/method$_\text{weak}$)-split($p\times l$o)/basis. Here, method$_\text{weak}$ and method$_\text{strong}$ stand for the theory level we employ to treat weakly and strongly correlated orbitals; $p$ is the number of groups and $l$o is the number of orbitals in the group; basis stands for the employed basis such as canonical molecular orbitals (CMOs), natural orbitals (NOs), or orthogonal atomic orbitals (SAOs).
The generalized version of SEET is denoted as SEET(method$_\text{strong}$/method$_\text{weak}$)-mix($l$o)/basis, where $l$o stands for the number of orbitals in each orbital group treated with accurate method; these orbital groups are intersecting.

Since different number of orbitals can be used to form groups of intersecting orbitals, we can create a series of approximations to the accurate SEET(method$_\text{strong}$/method$_\text{weak}$)-mix($N$o) which would include all $N$ strongly correlated orbitals in the subset treated by a high level, accurate method.
SEET-mix($2$o), SEET-mix($3$o), SEET-mix($4$o), etc., where we calculate the self-energy between any 2, 3, and 4 orbitals accurately using method$_\text{strong}$ can be evaluated as approximations to the accurate SEET-mix($N$o). In contrast to recovering only a subset of self-energy elements in SEET-split version, SEET-mix($l$o) always leads to recovering all the self-energy elements between $N$ strongly correlated orbitals with the accurate treatment of self-energy between $lo$ orbitals.
Such a treatment provides a smooth convergence of the energies to the SEET-mix($N$o) answer. 

\begin{figure} [h]
  \includegraphics[width=8.0cm,height=5.5cm]{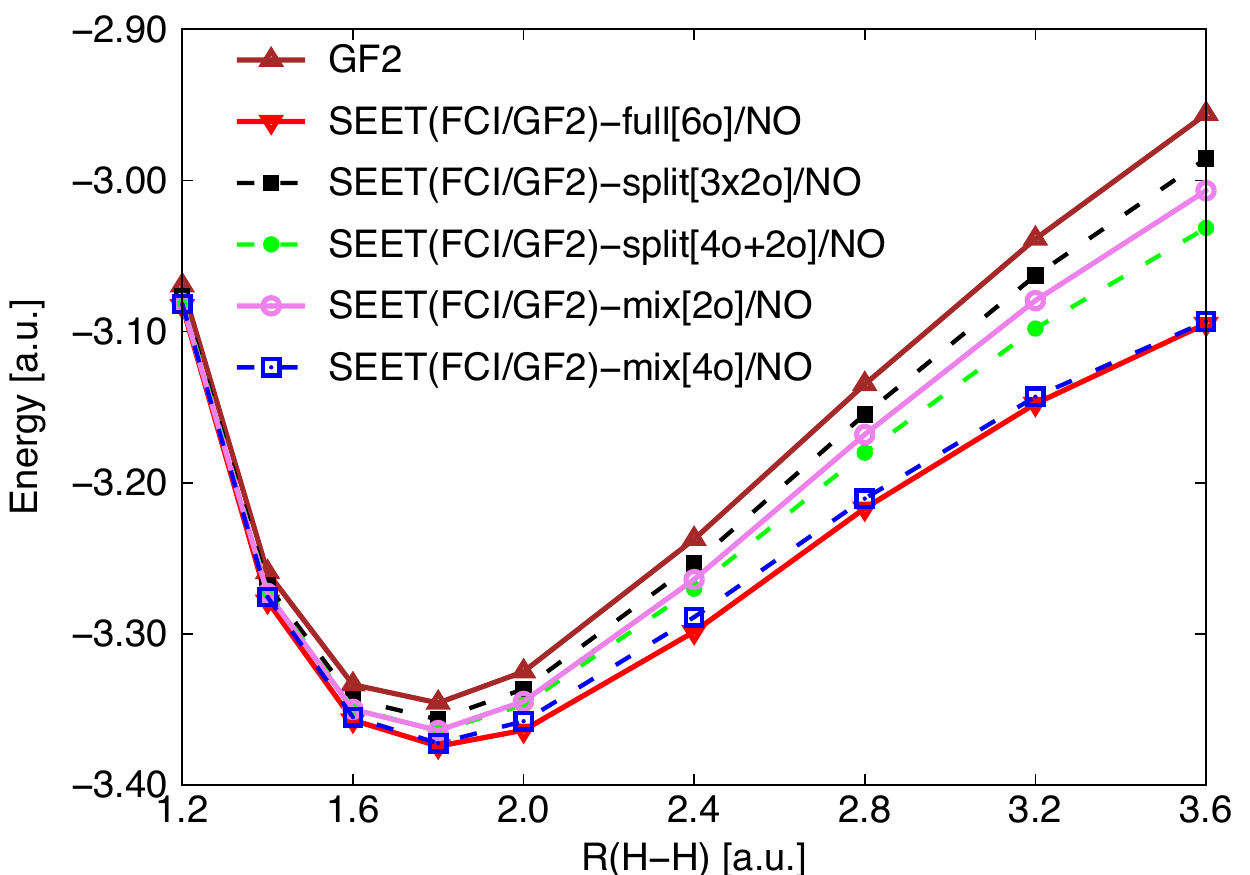}            
  \includegraphics[width=8.0cm,height=5.5cm]{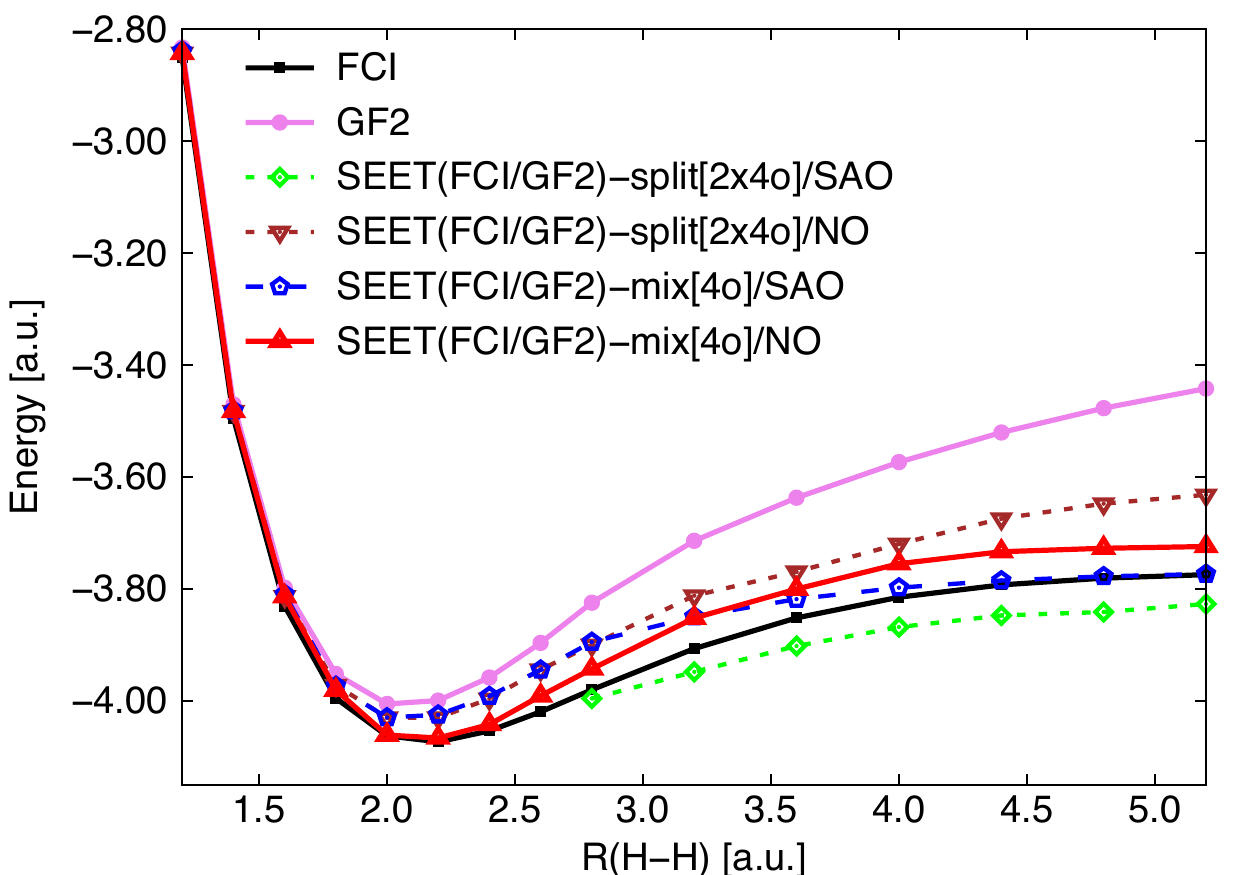}            
  \caption{\normalsize Upper panel: Potential energy curves for the H$_6$ chain in cc-pVDZ basis.
    Lower panel: Potential energy curves for the 2$\times$4 H lattice in STO-6G basis.}
  \label{fig:h6_h8}
\end{figure}

\begin{figure} [h]
  \includegraphics[width=10.0cm,height=9cm]{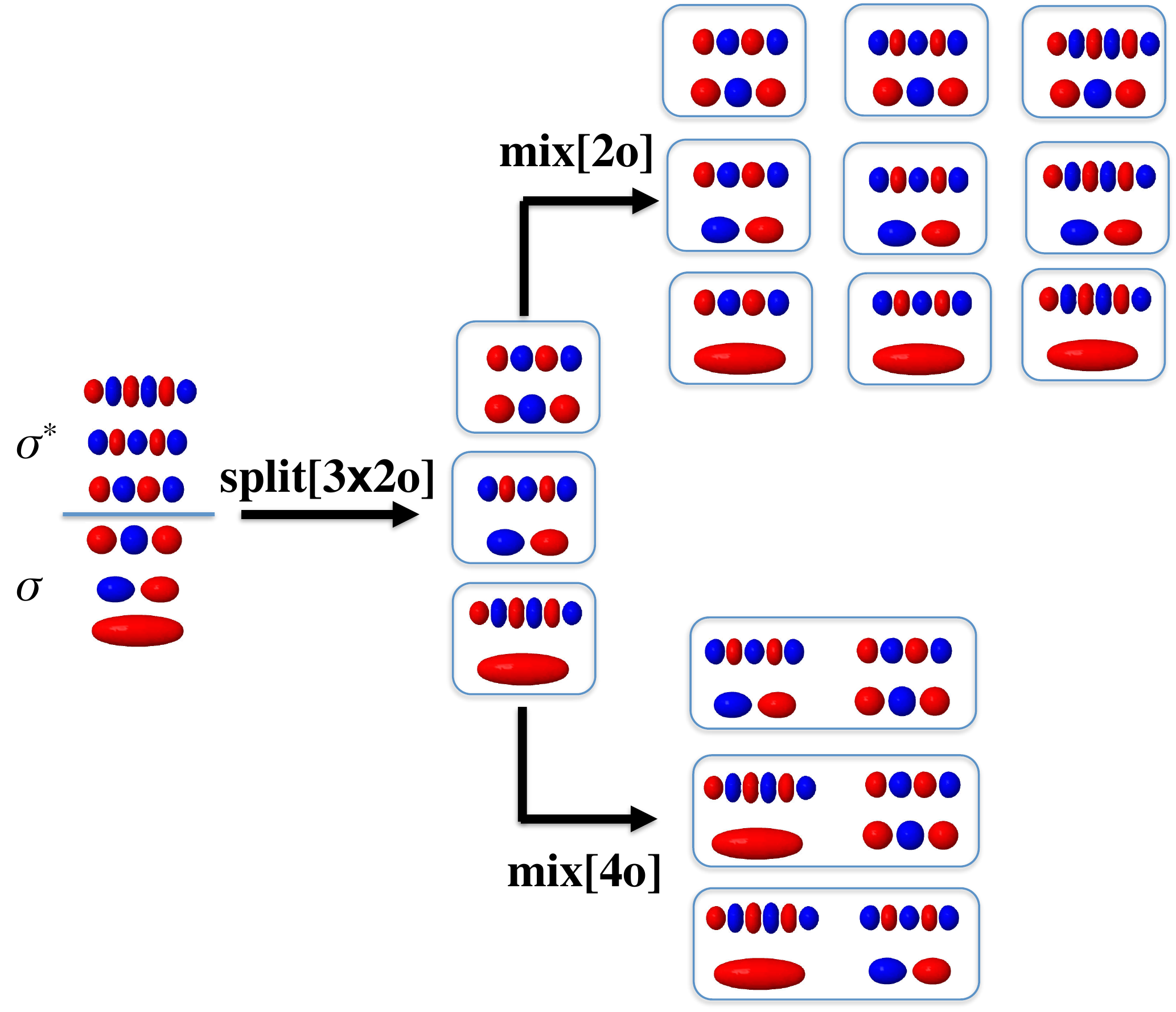}
  \caption{\normalsize Three $\sigma$ and three $\sigma^*$ orbitals are visualized to explain  split[3$\times$2o], mix[2o], and mix[4o] schemes for the H$_6$ chain.}
  \label{fig:s1}
\end{figure}

\begin{figure} [h]
  \includegraphics[width=9cm,height=6cm]{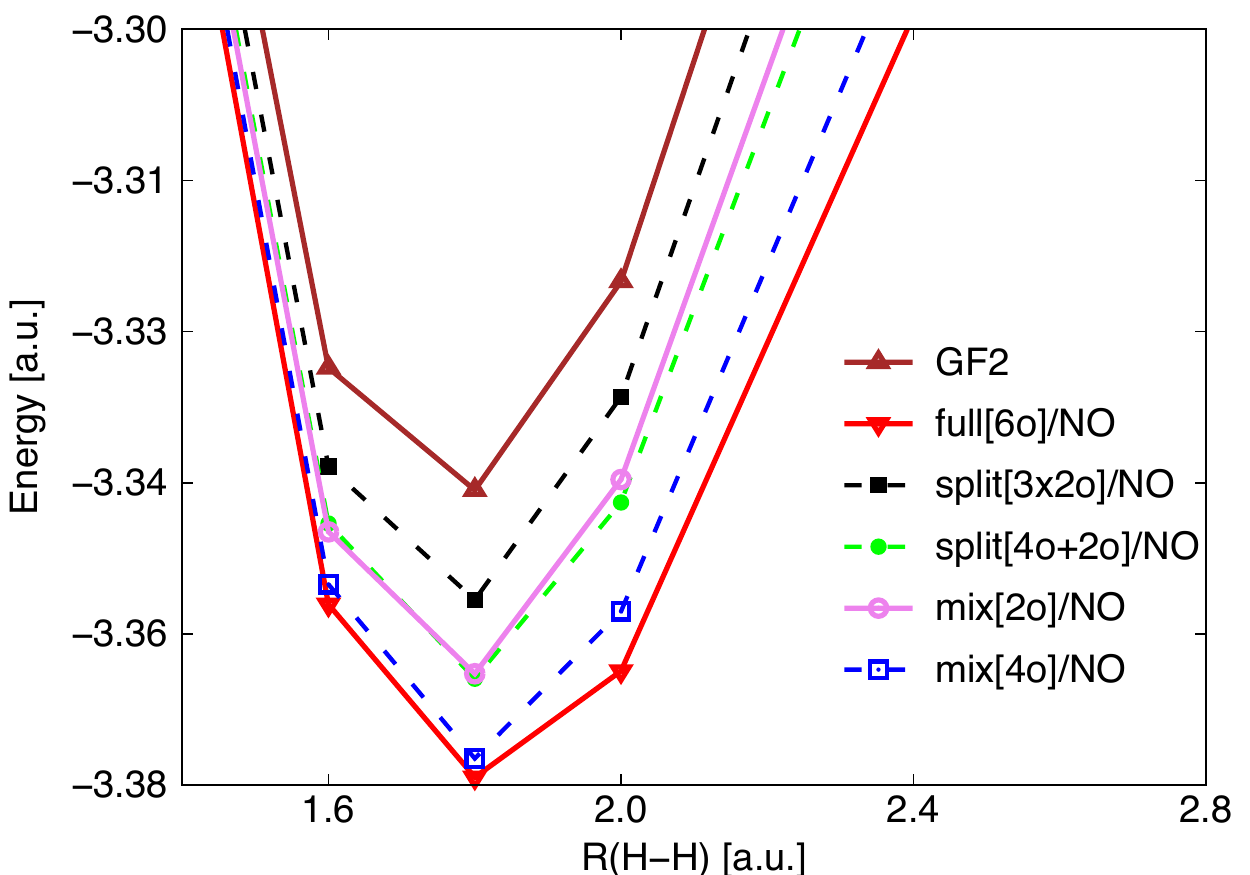}
  \caption{\normalsize Potential energy curves of H$_6$ chain in cc-pVDZ basis around the equilibrium geometry. With the exception of GF2, all other results are evaluated using  SEET(FCI/GF2) level of theory applying both the SEET-split and SEET-mix schemes.}
  \label{fig:s2}
\end{figure}

To demonstrate the potential of the generalized SEET, first we consider a simple H$_6$ chain and 2$\times$4 H lattice. The potential energy curves of these systems are shown in Fig.~\ref{fig:h6_h8}.
For the H$_6$ chain, all SEET calculations are performed in NOs and the SEET(FCI/GF2)-full[6o] energy is used as a reference.
We first split the full active space of 6 $\sigma-$type orbitals into three groups of bonding--anti-bonding pairs. This scheme, denoted as split[3$\times$2o], yields
energies  lower than GF2 but remains very far from the reference energy.
To improve the split[3$\times$2o] energy while keeping only two strongly correlated orbitals in each group, we allow each bonding orbital to couple with every of the anti-bonding orbitals resulting in mix[2o] scheme. The mix[2o] pairs are visualized in Fig.~\ref{fig:s1} of SI.
We observe that mix[2o] gives lower energies than the split[3$\times$2o] scheme.
Especially at the short distances, the mix[2o] energies are comparable to those of the split[4o+2o] (for zoom of the equilibrium geometry, see Fig.~\ref{fig:s2} of SI).
The mix[2o] scheme, however, is still insufficient to recover the full[6o] energy in the stretched regime. We therefore enlarge the active space by introducing the mix[4o] scheme.
While there are some possibilities to construct the active spaces of 4o from 6o using the mixing scheme, here we employ the most straightforward way where
 we simply mix the bonding--anti-bonding pairs which were used in the split[3$\times$2o] scheme (for details see Fig.~\ref{fig:s1} of SI).
The energies from mix[4o] approach remain close the full[6o] reference with errors of few mHa.

\begin{figure} [h]
  \includegraphics[width=7.0cm,height=13cm]{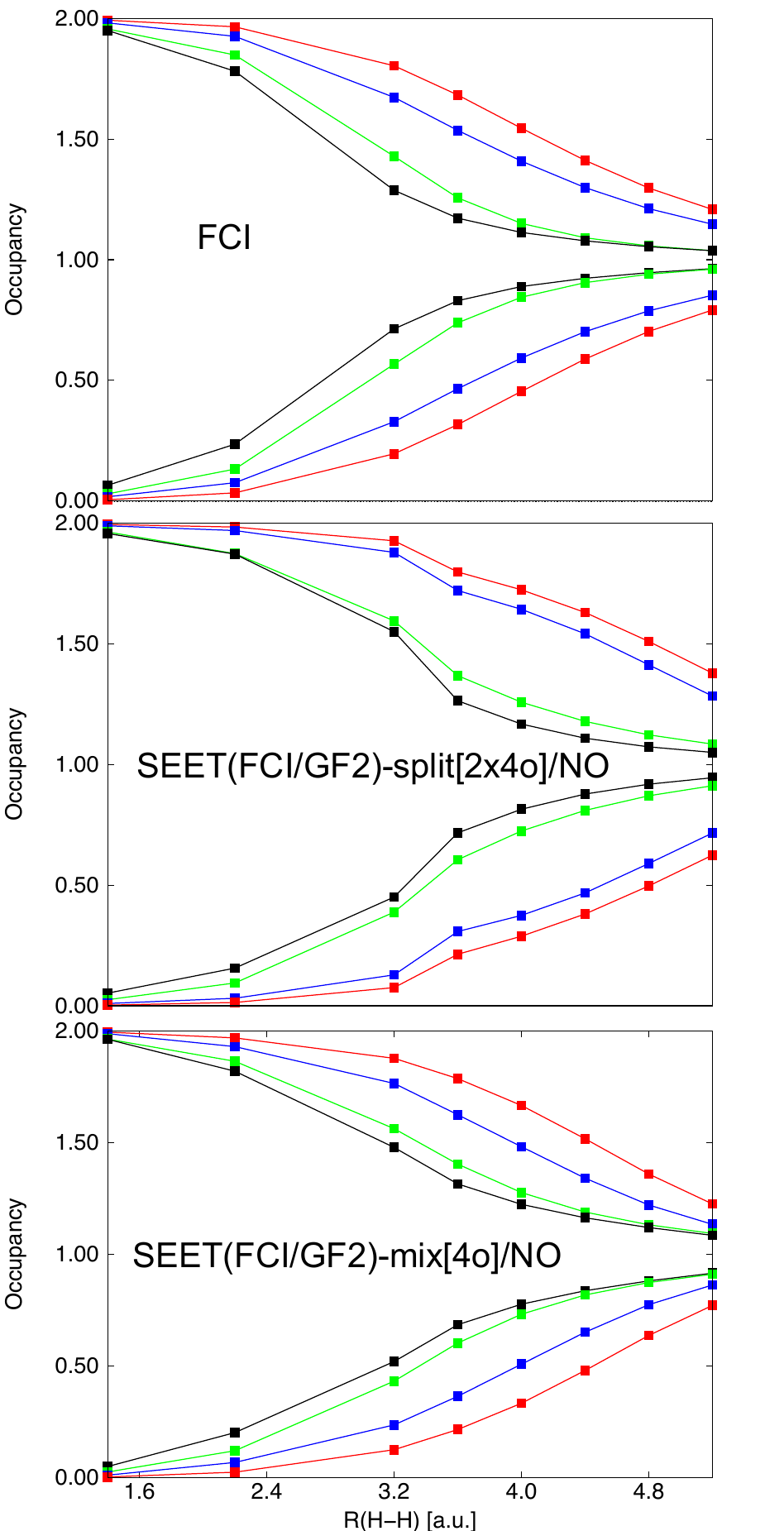}
  \caption{\normalsize Occupation numbers for the 2$\times$4 H lattice from FCI, SEET(FCI/GF2)-split[2$\times$4o]/NO, and SEET(FCI/GF2)-mix[4o]/NO calculations.}
  \label{fig:s3}
\end{figure}

Let us now consider the 2$\times$4 H lattice which is a more challenging example than the H$_6$ chain.
Here, we use FCI energy as a reference.
The split[2$\times$4o] calculation in NOs improves over GF2 but it is far from FCI due to the lack of correlations between strongly correlated orbital groups.
The split[2$\times$4o] scheme in SAOs can only converge at long distances and its curve goes below the FCI one because of the overcorrelation of GF2 present at long distances.
This indicates that the mixing scheme has to be introduced to recover the correlations between strongly correlated orbital groups in both bases.
The significant improvement upon the split[2$\times$4o] results is achieved for both bases in the mix[4o] scheme.
At short distances (R $<$ 2.4 a.u.), mix[4o] in NOs gives energies comparable to FCI. At long distances (R $>$ 3.4 a.u.) the curve of mix[4o] in SAOs is almost identical to the FCI one.
Additionally, we evaluate occupation numbers presented in Fig.~\ref{fig:s3}. We observe that the transition from single-reference to multi-multireference regime is smoother for the mixing than for splitting scheme, indicating that mix[4o] in NOs can correctly captures the static correlation.
This is reflected at long distances by the parallelity of the mix[4o]/NO curve to the FCI one.
\begin{figure} [h]
  \includegraphics[width=8.0cm,height=5.5cm]{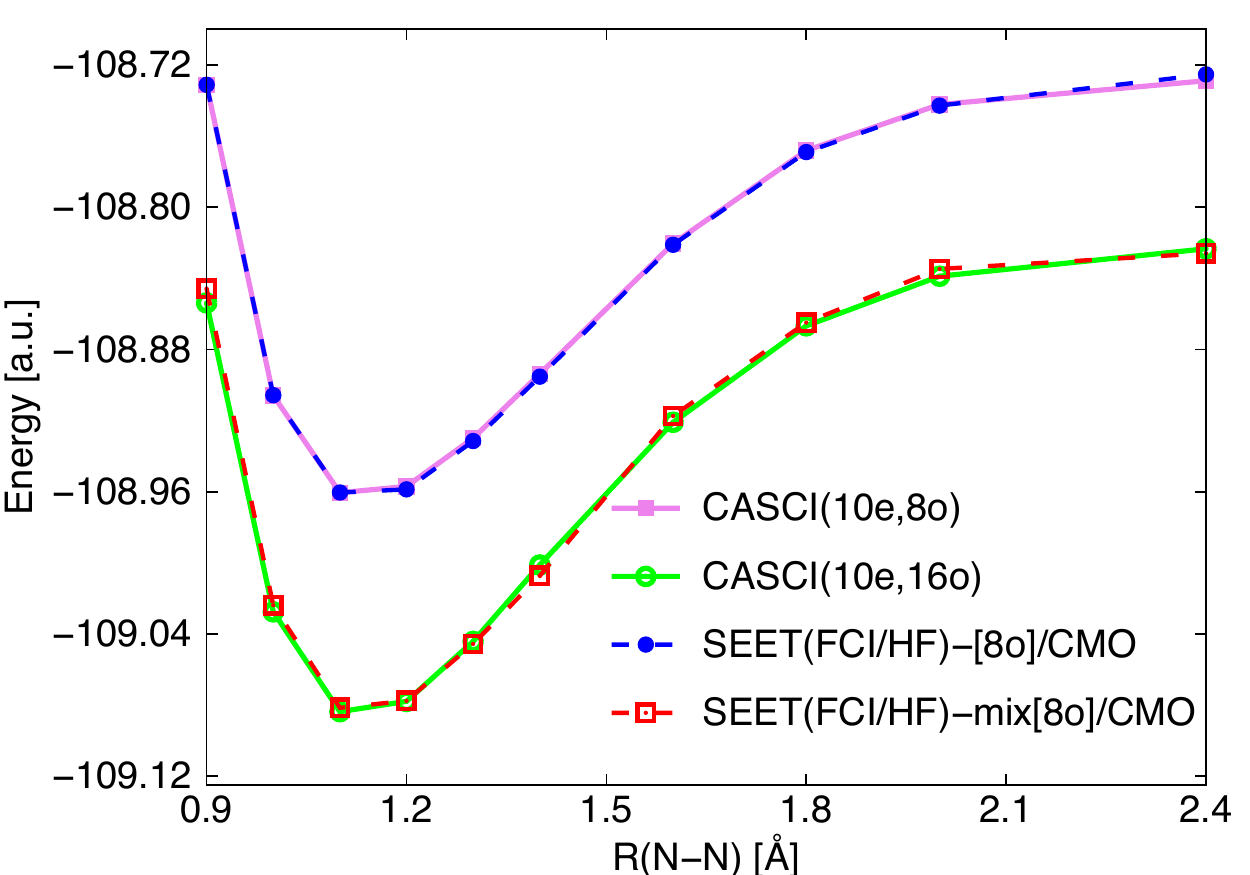}
  \caption{\normalsize Potential energy curves for N$_2$ in 6-31G basis.}
  \label{fig:n2}
\end{figure}

After assessments of the hydrogen clusters, we move on to more difficult cases.
Stretching the triple bond of N$_2$ is a difficult test case for many quantum chemistry methods and consequently 
an interesting case for assessing the mixing scheme.
In order to fully explore the performance of mixing scheme, the SEET(FCI/HF) calculation in HF CMOs was used.
Fig.~\ref{fig:n2} displays potential energy curves of N$_2$ in the 6-31G basis.
We provide two CASCI calculations, CASCI(10e,8o) and CASCI(10e,16o) as points of comparison to the SEET results.
The active space consists of 8 valence orbitals in CASCI(10e,8o), while CASCI(10e,16o) involves all 16 orbitals without considering 1$s$ orbitals.
As expected, the SEET(FCI/HF)-[8o] curve, where all the active orbitals are placed in one group, coincides with the CASCI(10o,8o) one.
It is interesting if the CASCI(10e,16o) curve can be reproduced using mixing scheme with fewer than 16 orbitals in the group.
To construct the orbital groups needed for the mixing scheme, we first divide the 16 orbitals into four groups of four orbitals according to types of orbitals: $s$, $p_x$, $p_y$, and $p_z$.
The groups of 8 strongly correlated orbitals were then constructed from these groups of four orbitals using mixing scheme.
As seen from Fig.~\ref{fig:n2}, the SEET(FCI/HF)-mix[8o] calculation excellently reproduces the CASCI(10e,16o) result.
\begin{figure} [h]
  \includegraphics[width=7.0cm,height=7.0cm]{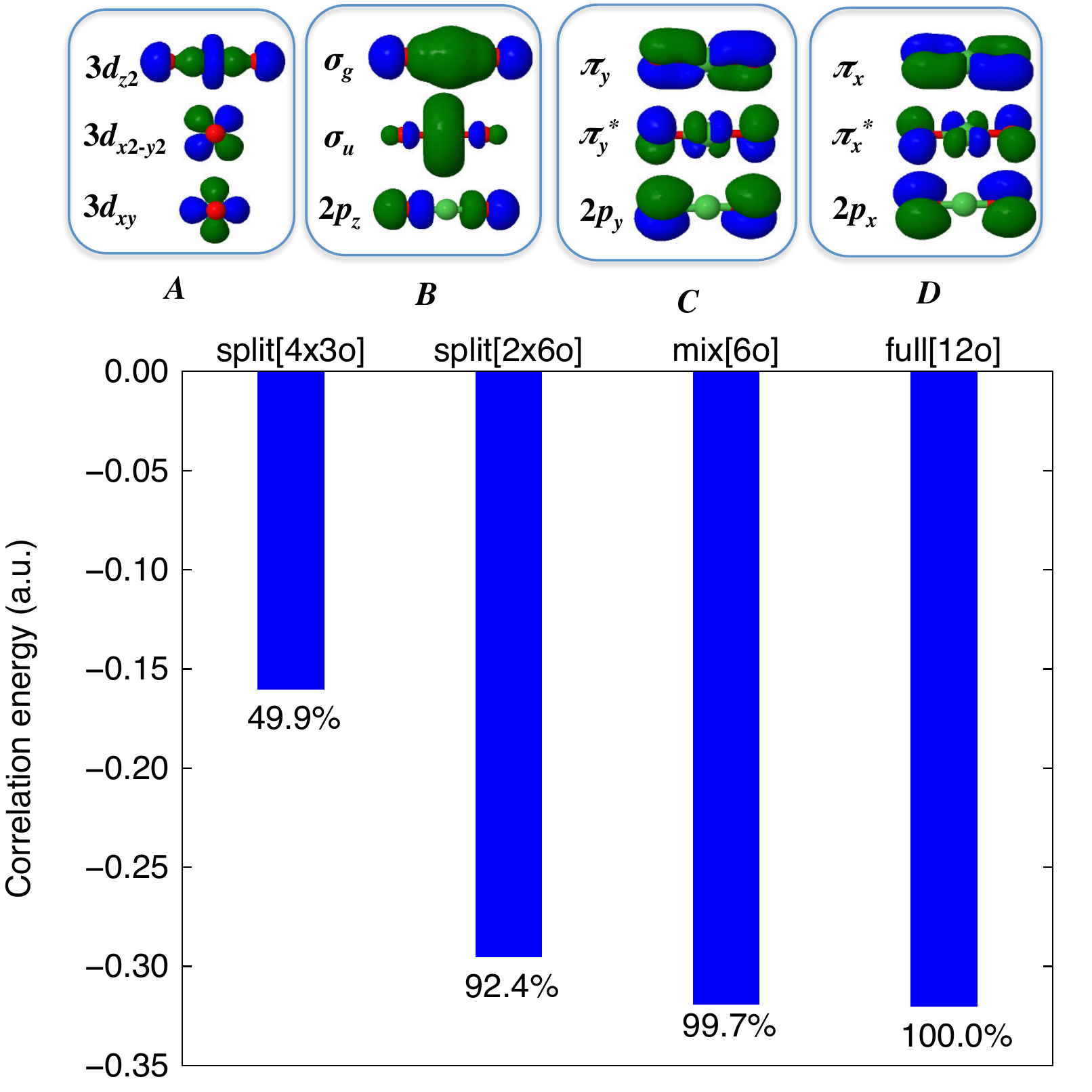} 
  \caption{\normalsize Upper panel: Four groups of valence orbitals in NiO$_2$.
    Lower panel: 
    The percentage  of recovered correlation energy in various SEET calculations in comparison to SEET(FCI/HF)-full[12o] in cc-pVDZ basis.}
   \label{fig:nio2}
\end{figure}

\begin{table}[!h]
  \normalsize
  \caption{\label{tab:nio2} \normalsize Occupancies of $\pi$--type orbitals in NiO$_2$.}
  \begin{tabular}{cccccccccc}
    \hline \hline
     Orbitals       &split[3$\times$4o] &split[2$\times$6o]  &mix[6o] &CASCI(18e,12o) \\
    \cline{1-5}
    $\pi_x$         &1.95               &1.77                &1.77    &1.78           \\
    $\pi_y$         &1.95               &1.77                &1.77    &1.78           \\
    $\pi^*_x$       &0.24               &0.45                &0.47    &0.47           \\
    $\pi^*_y$       &0.24               &0.45                &0.47    &0.47           \\
    \hline \hline
  \end{tabular}
\end{table}

The next system used to investigate the performance of SEET-mix is a linear NiO$_2$ molecule in cc-pVDZ basis,
which is a challenging system because of many low-lying states are present close to the ground state~\cite{hubner2012cyclic,vogiatzis2015systematic}.
We first divide the full active space including six $3d4s$ orbitals of Ni atom and six $2p$ orbitals of two O atoms into four groups $A, B, C, D$ displayed in the upper panel of Fig.~\ref{fig:nio2}.
The strongly correlated orbital groups are constructed as follows in different SEET schemes: split[4$\times$3o] = $\{A, B, C, D\}$, split[2$\times$6o] = $\{A+B, C+D\}$, and mix[6o] = $\{A+B, B+C, B+D,C+D\}$. 
We employed the SEET(FCI/HF) calculation, as opposed to the SEET(FCI/GF2) one, because we only aim to focus on the strong correlations within the active space.
To check if SEET-mix and SEET-split can correctly describe the ground state of the linear NiO$_2$ molecule, we are comparing the occupation numbers from SEET calculations with the CASCI(18e,12o) calculation. Those for $\pi-$type orbitals are summarized in Table~\ref{tab:nio2}.
SEET-split[4$\times$3o] provides incorrect occupation numbers.
For the enlarged orbital space of 6 orbitals, i.e. SEET-split[2$\times$6o] and SEET-mix[6o], the occupation numbers are correctly recovered and comparable to those from CASCI(18e,12o). This means that both split[2$\times$6o] and mix[6o] are correctly describing the ground state of NiO$_2$.
The correlation energies from SEET(FCI/HF) calculation are presented in the lower panel of Fig.~\ref{fig:nio2}. 
Here, for internal consistency, the SEET(FCI/HF)-full[12o] correlation energy is used as a reference.
Although SEET-split[2$\times$6o] gives similar occupation numbers to  SEET-mix[6o], it cannot fully recover the correlation energy due to missing of correlations between two orbital groups contributing to the self-energy. SEET-mix[6o] recovers the correlation energy within the full active space up to 99.7\%.  

\begin{figure} 
 \includegraphics[width=8.0cm,height=5.5cm]{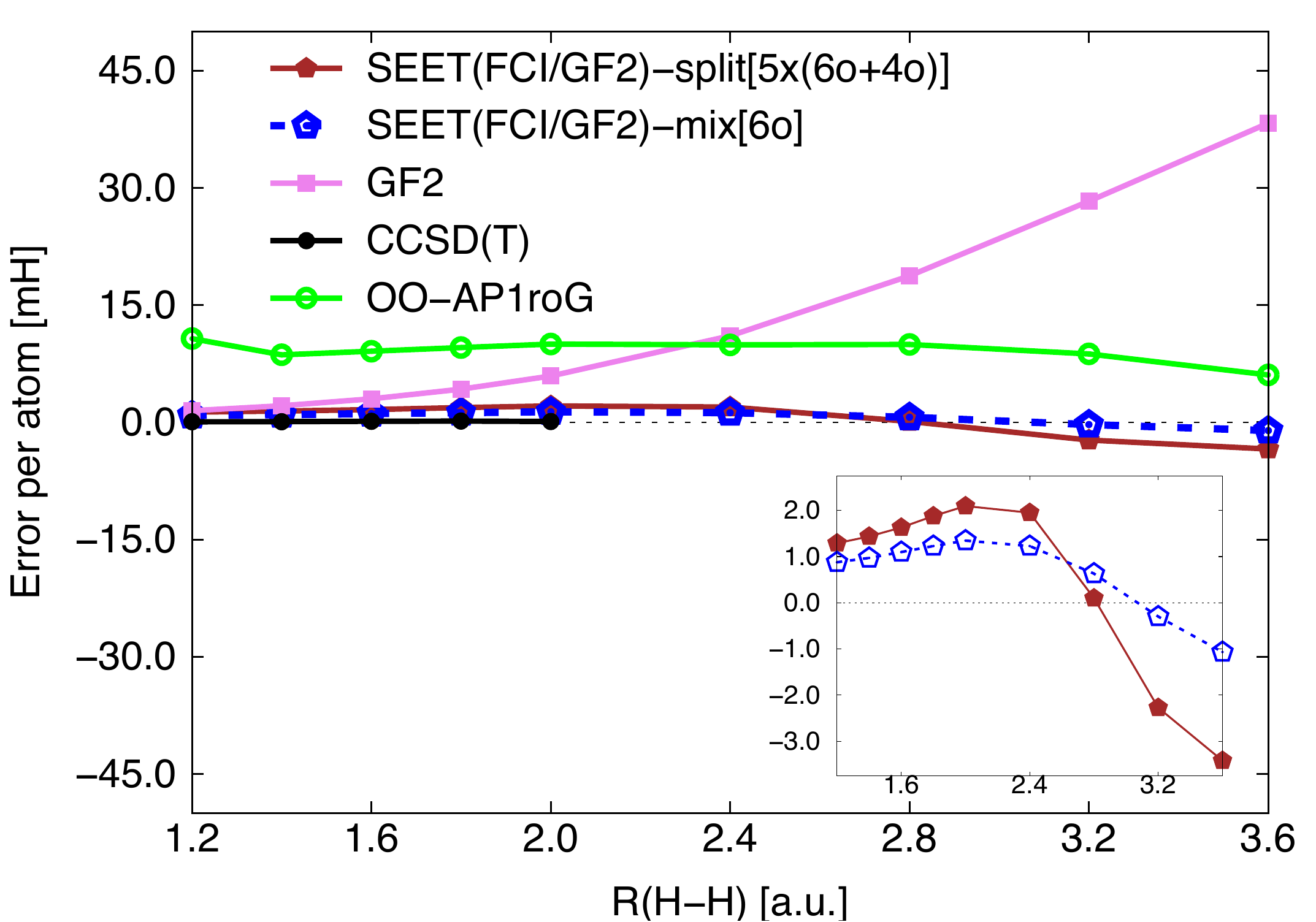}     
 \caption{\normalsize Energy error per atom (in mHa) with respect to the DMRG reference from various methods for the H$_{50}$ chain in STO-6G basis. The DMRG and OO-AP1roG data are taken from Refs.~\citenum{Hachmann:jcp2006-h50dmrg,Boguslawski2014}.}
  \label{fig:h50}
\end{figure}

\begin{figure} [h]
  \includegraphics[width=9.0cm,height=6cm]{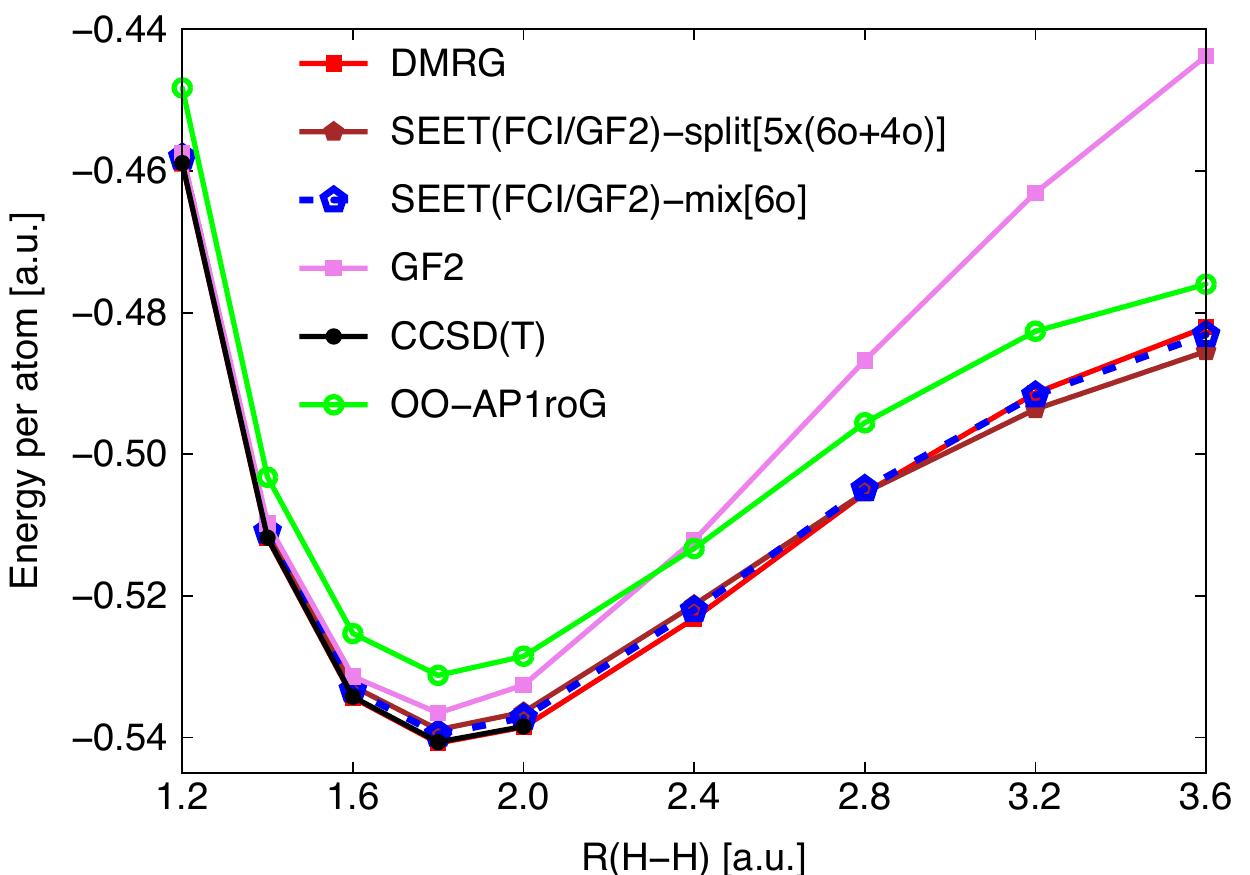}
  \caption{\normalsize Potential energy curves of H$_{50}$ chain in STO-6G basis.The DMRG and OO-AP1roG data are taken from Refs.~\citenum{Hachmann:jcp2006-h50dmrg,Boguslawski2014}.}
  \label{fig:s4}
\end{figure}

Finally, we examine the generalized SEET scheme on the H$_{50}$ chain in the STO-6G basis. This is a well-known benchmark for strongly correlated methods since the active space is large and contains 50 electrons in 50 orbitals. Note that for this system traditional single reference methods such as CCSD(T) are unable to converge past the distance of 2 a.u. The reference solution is available from DMRG calculations~\cite{Hachmann:jcp2006-h50dmrg}. 
Although SEET-split results are closer to the DMRG reference than the other method capable of targeting for strong correlations such as orbital-optimized antisymmetric product of one-reference-orbital germinal (OO-AP1roG)~\cite{Boguslawski2014}, the errors of SEET-split are still present, especially at long distances. For long distances, SEET-mix[6o] yields a significant improvement over SEET-split scheme as demonstrated in errors with respect to the DMRG reference shown in Fig.~\ref{fig:h50}.
The potential energy curves are shown in Fig.~\ref{fig:s4} of SI. 

We conclude that the generalized SEET, denoted as SEET-mix, yields quantitatively accurate results for cases where active space containing $N$ strongly correlated orbitals is separated into a series of intersecting smaller groups/active spaces of $K$ strongly correlated orbitals, where $K<N$.
SEET-mix can be employed to create a hierarchy of systematic approximations, as a function of $K$, to the exact Luttinger-Ward functional and the resulting self-energy. In such a hierarchy, the self-energy elements between $K$ strongly correlated orbitals are recovered by an accurate method or in other words any possible occupation of the group of $K$ orbitals is explored in the presence of many-body field coming from all the other orbitals~\cite{Rusakov14}. Thus, such a scheme has some similarities with the coupled clusters (CC)~\cite{Cizek_CC,Paldus_Cizek} or method of increments~\cite{Paulus20061,Stoll200990} hierarchy. However, while some similarities exists, in stark contrast to the standard CC and method of increments, SEET-mix does not result in divergences and qualitatively incorrect answers when less than full number of active orbitals (or excitations) is used when strong correlations are present. 

Let us also note that SEET-mix shares some similarities with the DMET bootstrap embedding procedure~\cite{DMET_bootstrap_jcp_2016,DMET_bootstrap_molphys_2017} in which the elements of the density matrix are produced by considering a series of spatially overlapping fragments. Similarly to  DMET bootstrap, SEET-mix can be performed using spatial orbitals, however, it also can be performed in the energy basis (NOs or MOs)  with more abstract criteria for forming intersecting orbital groups. The energy basis is commonly employed for molecular problems by us since it is advantageous to initially separate weakly and strongly correlated orbitals.

Moreover, in contrast to now standard quantum chemistry methods for treating large active spaces such as density matrix renormalization group (DMRG)~\cite{White_DMRG_1992,Chan_dmrg_2002,Legeza_dmrg_2003,Sharma_Chan_review_2011}, SEET-mix does not suffer from limitations such as orbital ordering present during the DMRG sweep procedure)~\cite{Reiher_ordering_2005}. 

Finally, SEET-mix provides a good framework for generalization to large active spaces that are notoriously difficult for standard quantum chemistry methods since all the intersecting groups of orbitals can treated simultaneously in an embarrassingly parallel fashion.

D.Z. acknowledges support from the a National Science Foundation (NSF) grant No. CHE-1453894. T.N.L. acknowledges support from the Simons Collaboration on the Many-Electron Problem.

\bibliographystyle{apsrev4-1}

\clearpage
\onecolumngrid
\section*{Supporting Information}

\end{document}